\documentstyle[aps, prb, epsf, multicol]{revtex}  
  
\begin{document}  
\title{Ground state and excitation of an asymmetric spin ladder model} 
\author{Shu Chen\footnote{
Present address: Institute for theoretical physics IV, 
Heinrich-Heine-Universit\"{a}t, D-40225 D\"{u}sseldorf, Germany}, 
H. B\"{u}ttner, and J. Voit}  
\address{  
Theoretische Physik 1, Universit\"{a}t Bayreuth, D-95440 Bayreuth, Germany}  
\date{\today}  
\maketitle  
  
\begin{abstract}  
We perform a systematic investigation on an asymmetric zig-zag spin ladder 
with inter-leg exchange $J_1$ and different exchange integrals $J_2 \pm
\delta$ on both legs. In the weak frustration limit, the spin model can be
mapped to a revised double frequency  sine-Gorden model by using bosonization.
Renormalization group analysis shows that the Heisenberg critical point flows
to an intermediate-coupling fixed  point with gapless excitations and a
vanishing spin velocity. When the frustration is
large, a spin gap opens and a dimer ground state is realized. Fixing $J_2 = J_1
/2$, we find, as a function of $\delta$, a continuous manifold of Hamiltonians
with dimer product ground states, interpolating between the  Majumdar-Ghosh and
sawtooth spin-chain model. While the ground state is  independent of the
alternating next-nearest-neighbor exchange $\delta$, the gap  size of
excitations is found to decrease with increasing $\delta$.  We also  extend
our study to a two-dimensional double layer model with an exactly known ground
state. \\    

\end{abstract}  
 
\begin{multicols}{2}  
\narrowtext  
\section{Introduction} 
Quantum spin ladder systems have attracted much attention in the past 
few years\cite{Dagotto,Dagotto2}. Strong quantum fluctuations prevent any long range 
antiferromagnetic (AF) order in quasi-one dimension. The magnetic 
phases of the ladder systems are rich and strongly dependent on their 
geometric structures. Several types of disordered ``quantum spin liquid'' phases 
are known \cite{Dagotto,Nersesyan,Wang}. Typical examples of two-leg ladders 
are the railroad ladder and the zigzag ladder. The railroad ladder has a
singlet  ground state with elementary triplet excitations (magnons)
\cite{Dagotto}.  Depending on the ratio of the leg to rung exchange integrals,
the zig-zag  ladders may have gapless ground states with algebraically
decaying spin  correlations or spontaneously broken dimerized ground
states\cite{Haldane}. The  gapped dimer ground state is degenerate, and the
elementary excitations are pairs of spinons. The spin ladders have been studied 
experimentally in compounds such as $SrCu_2O_3$ and $CuGeO_3$ \cite{Castilla}. 
 
\begin{figure}[t]  
\centerline{\epsfysize=2.5cm \epsffile{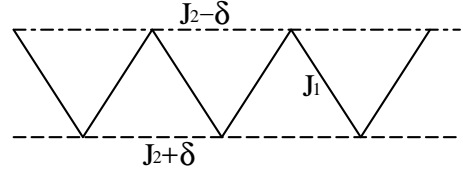}}  
\caption{The asymmetric zig-zag spin ladder with different exchanges on
the top and bottom legs.}   
\label{mod_fig}  
\end{figure}

The two-leg zigzag ladder, which has been well investigated, is perhaps the simplest 
example of the frustrated spin model and highlights the role played by 
frustration. However, less attention has been paid on asymmetric spin 
ladders where the exchange integrals on both legs are different. Only the extreme 
case where one  leg of a zig-zag ladder is missing entirely (sawtooth or 
$\Delta$-chain) has  been solved\cite{Sen,Nakamura}. In this paper, we perform 
a systematic study of an asymmetric zig-zag spin ladder, which is a Heisenberg 
model defined on the structure shown in Fig. \ref{mod_fig}. In general, it  
is convenient to represent the zigzag ladder as a spin chain with
nearest-neighbor (NN) exchange $J_1$ and next-nearest-neighbor (NNN) exchange
$J_2$ corresponding to the inter-leg exchange $J_1$ and intra-leg exchange
$J_2$. Equivalently, the asymmetric ladder model can be represented as a
chain with  an alternating NNN exchange   
\begin{equation}   
H=\sum_{l} \left\{ J_{1}{\bf S}_{l} \cdot 
{\bf S}_{l+1}+  \left[ J_{2}+(-1)^{l}\delta \right]  {\bf S}_{l} \cdot {\bf 
S}_{l+2} \right\} \; ,  
\label{spinchain}   
\end{equation}   
where $J_{1}\equiv 1$ and $J_{2}\pm \delta$ are the nearest-neighbor (NN)   
and alternating NNN coupling constants, respectively. The introduction of $\delta$ 
makes the exchanges on top and bottom legs different. $\delta=0$ 
is the ordinary zig-zag ladder  or frustrated spin chain\cite{Haldane} and  
$\delta = J_{2}$ is the extreme case with one leg completely missing. 
Fixing $J_{2}=J_{1} /2$ gives an exactly solved continuous manifold 
of Hamiltonians \cite{CBV}. The Majumdar-Ghosh (M-G) \cite{M-G} and 
the sawtooth chain \cite{Sen,Nakamura} are extreme cases with  $\delta 
= 0$ and $\delta = J_{2}= J_{1} /2 $ respectively. 
       
For $\delta = 0$ the model (\ref{spinchain}) is well understood for general
$J_2$ \cite{Haldane}. Frustration due to $J_2$ is irrelevant when $J_2 <
J_{2c}$ \cite{Castilla,Okamto}, and the system renormalizes to the Heisenberg fixed
point \cite{Affleck}, whose ground state is described as a spin fluid or 
Luttinger liquid with massless spinon excitations. As $J_2 > J_{2c}$, the 
frustration term is relevant and the ground state is doubly degenerate. 
Particularly, the ground state has 
simple product form at $J_{2}=0.5 J_{1}$\cite{M-G}. The critical value of 
${J_{2c}}=0.2412 J_{1}$ can be determined numerically\cite{Castilla,Okamto}. For 
$J_{2}>0.5J_{1}$, quite different  field theory treatments are required 
depending on the ratio of  ${J_{2}}/{J_{1}}$ \cite{Allen2,White}.  
   
Since relatively less is known about the asymmetric ladder, we 
study physical effects brought about by the 
leg-asymmetry. 
Before solving the quantum problem, we start with the classical problem
which can give us an instructive insight into properties of the asymmetric
spin model. In the classical limit, the ground state of this model is a
N\'{e}el state for  $J_2 < J_1/4$ and a spiral with a pitch $\alpha =
\arccos(-J_1/4J_2)$  for  $J_2 > J_1/4$. Both ground states and the critical
ratio of exchange integrals separating them, are independent of $\delta$. The
excitations may depend on  $\delta$, however.    

Certainly, the quantum case is much more complicated. When the asymmetric 
exchange interaction is introduced, some unexpected phenomena will appear. We 
found that the Heisenberg fixed point is no longer stable and flows to an 
intermediate-coupling fixed point with gapless excitations and a vanishing 
spin velocity. \cite{CBV} However, there is still reminiscence of 
classical results. It was found that the ground state
is independent of $\delta$ when $J_2 =  J_1/2$, but the excited gap is
decreased by $\delta$. Part of the work on weak frustration regime has 
been reported in our previous letter\cite{CBV}, and here we quantitatively study 
the crossover of the excitation spectrum from the symmetric M-G model to the extremely 
asymmetric sawtooth model in detail. An extension to a two-dimensional double 
layer model is also presented.        

The outline of this paper is as follows. In section II, the effective low 
energy theory of the asymmetric ladder is derived by using bosonization. We 
qualitatively discuss the effect of alternating NNN interaction and compare 
our model with the well-known spin-Peierls model. The phase diagram of the 
system is discussed with the help of the renormalization group analysis. In section
III,  we study the asymmetric model at the special point $J_2 =
J_1/2$. The  ground state and excitation properties are also discussed. In
section IV, we  generalize our model to a two-dimensional double layer model,
whose ground  state is a dimer product state and the excitations are magnons.
Section V contains our conclusions. 
 
\section{Effective low-energy theory and Renormalization group analysis} 
Following the general procedure of transforming a spin model to an effective model of 
continuum field \cite{Luther,Affleck},  we convert the spin Hamiltonian to a 
Hamiltonian of spinless fermions using Jordan-Wigner transformation, then map
it to a modified Luttinger model with Umklapp and backscattering-type interactions. 
Using the standard dictionary of bosonization\cite{Voit1,Delft}, we obtain the 
effective boson 
Hamiltonian $H = H_0 + H_1$ with
\begin{eqnarray}   
H_{0} & = & \int dx\frac{u}{2\pi } \left[ K(\pi \Pi
)^{2}+\frac{1}{K}(\partial_{x}\Phi )^{2}\right] \; ,  
\label{h0} \\   
H_1 & = & \int dx \left[    
{\frac{g_{3}}{2(\pi a)^{2}}}\cos 4\Phi+{\frac{g_{1}}{\pi ^{2}a}}   
\left( \partial_{x} {\Phi} \right) \cos 2\Phi   
\right] \; , 
\label{h1}   
\end{eqnarray}   
where $\Phi(x) $ is a bosonic phase field and $\Pi(x) $ its canonically conjugate 
momentum. Here $a$ is a short-distance cutoff, $g_{3} \propto 1- J_{2}/J_{2c}$ is the 
Umklapp-scattering amplitude and $g_{1} \propto \delta$ is the amplitude 
of the alternating NNN field. 
The parameters $u$ and $K$ are the effective spin velocity and coupling constants
which are given by  
\[ 
u=\sqrt{(1+\frac{g_{4}}{{2\pi}})^2-(\frac{g_2}{{2\pi}})^2},\ \ \ \ 
 K=\sqrt{ 
\frac{{2\pi+g_{4}-g_{2}}}{{2\pi+g_{4}+g_{2}}}}.  
\] 
In genereal, these values are only valid near the free fermion point ($K=1$),
whereas $K = 1/2$ is fixed by the symmetry at the isotropic point. The 
corresponding spin-correlation functions 
can be calculated from the boson representation, which gives   
\begin{eqnarray}    
\left\langle S^{z}(0)S^{z}(x)\right\rangle _{0}\  &\sim &\ 
(-1)^{x}x^{-2K},  \\  
\left\langle S^{+}(0)S^{-}(x)\right\rangle _{0}\  &\sim &\ (-1)^{x}  
x^{-\frac{1}{2K}}.  
\end{eqnarray}  
It is clear that the SU(2) symmetry is restored at the isotropic point with 
$K = 1/2$.  
 
For small $g_3$ and $g_1$, $H_{1}$ could be considered as a perturbation
to $H_0$. Without the $g_1$ term, $H = H_{0} + H_{1}$ represents a
standard  sine-Gorden model \cite{Haldane}. The $g_3$ term is either
marginally  irrelevant, which leads to the weak-coupling Heisenberg fixed
point, or relevant,  which drives the system to a strong-coupling dimer state.
Therefore, what we are interested in is how the alternating NNN interaction
$g_1$ changes the physical properties of the
system. Qualitative results on the influence of the new interaction ($g_1$)
can be obtained from scaling analysis and physical considerations. 
 
An important information about perturbative operators is whether they  
are relevant, marginal or irrelevant. In general, only the most relevant 
perturbation is important, because the irrelevant operator will scale to zero at 
large lengths. We can give an approximate estimate by comparing
scaling dimensions of the given operators. It follows that $e^{i \alpha  
\Phi(x)}$ has scaling dimension of $\frac{\alpha^2 K}{4}$ and $ \partial
_x\Phi   e^{i \alpha \Phi(x)}$ has scaling dimension of $\frac{\alpha^2
K}{4}+1$. Therefore, the scaling dimensions of the Umklapp and the
alternating NNN terms $g_3$ and $g_1$ are     
\begin{equation}  
\label{scaldim}  
d_{g_3}=4K, ~~~ d_{g_1}=K+1 .  
\end{equation}    
At the isotropic Heisenberg fixed point, $g_{3}$ is   
marginal with $d_{g_{3}}=2$, while the $g_{1}$-term with $d_{g_{1}}= 3/2$    
is relevant. We conclude that $g_1$ destabilizes the isotropic Heisenberg   
fixed point and the spin liquid ground state.  
 
On the other hand, for $J_2>J_{2c}$, the $g_3$ term is marginally relevant
and renormalizes to a strong coupling fixed point in the long-wavelength
limit. Near the strong coupling dimer fixed point, the $g_3$ term is much
more relevant than the $g_1$ term. Usually (e.g. $g_3 \rightarrow \pm
\infty$), the boson field $\Phi(x)$ locks into a constant value with small 
fluctuations, and an associated excitation gap. Here, the constant solution 
$\Phi = \pm  \pi/4$ corresponds to the degenerate ground state at the strong 
coupling fixed point for $g_{3} \rightarrow \infty$. The standard $\cos4\Phi$ 
sine-Gordon equation has a pair of solutions of kink and antikink, which 
describe the elementary excitations (a pair of spinon) for the degenerate dimer 
phase. Even including the less relevant $g_1$ term, the soliton solutions 
will survive. However, the phase locking of $g_1$ term is forbidden by the 
$\partial_x \Phi$-pre-factor to the $\cos(2 \Phi)$-term in $H_1$. In this 
sense, there is no standard strong coupling theory for the $g_1$-term. 
 
From known results on the sawtooth chain\cite{Sen,Nakamura} 
and the M-G model\cite{M-G}, we expect that the $g_1$ term, induced by the 
alternating NNN  interaction, does not confine spinons and plays a quite different
role than the dimerization by other degrees of freedom. Moreover,
the difference in the size of excitation gaps in these two models implies
that the $g_1$ term  quite generally competes with the Umklapp term whereas an
external NN  dimerization would cooperate. As we will show, it turns out that
$g_1$ opens no spin gap  despite being a relevant perturbation of the
Heisenberg fixed point. This result is also corroborated by the absence of a
magnetization plateau in our model in small magnetic fields\cite{Wiessner}.  
For an alternating NN exchange,  a
magnetization plateau is observed in small magnetic fields, but for alternating 
NNN exchange, it is only observed in high fields \cite{Wiessner,Fled}.    

To get an instructive insight, we would like to compare the alternating
NNN interaction in our model with the alternating NN interaction in the
well-known spin-Peierls (SP) model. In the language of field theory, the
external dimerization corresponds to a  relevant term
 $H_{sp} \propto \int dx  ( g_{sp} \sin 2\Phi) $. This
term is always much more relevant than the $g_3$ term and lifts the
degeneracy ($\Phi = \pm  \pi/4$) of the ground state. With the presence of the
$g_{sp}$ term, the lowest energy configuration is  $-\pi/4$ if $g_{sp}>0$ or
$\pi/4$ if $g_{sp}<0$. Since the most relevant term is $\sin 2\Phi$, the
corresponding sine-Gordon system has a pair of soliton and antisoliton
solutions (excitations with $S_z=\pm 1$) as well as two breather solutions
(excitations with $S_z=0$).\cite{Dashen,Tsvelik} The lowest breather is
precisely degenerate with the soliton and antisoliton excitations, and they
form a bound state which corresponds to $S=1$ triplet. In this case, the
elementary excitation should be a spin triplet and a spin singlet, no spinons
exist as elementary excitation.\cite{Affleck2,Dobry,Uhrig} However, the
alternating NNN  interaction, $\left( \partial_{x}  {\Phi} \right) \cos
2\Phi$, does not lift the degenerate phases $\Phi = \pm  \pi/4$, due to the
existence of the pre-factor $\partial_{x} {\Phi}$. 

We now perform a perturbative renormalization group (RG) analysis to our  
model by following standard procedures.\cite{chui,KT,Giamarchi,Voit2}
Introducing the reduced variables $y_{3}=\frac{g_{3}}{\pi u}$ and 
$y_{1}=\frac{g_{1}}{\sqrt{2}\pi u}$,  
we obtain the RG equations
\begin{eqnarray}   
\label{dk}   
\frac{dK}{dl}\, &=&\,-y_{3}^{2}{K}^{2}+y_{1}^{2}{K}^{4}\, \\   
\label{dy3}   
\frac{dy_{3}}{dl}\, &=&\,(2-4K)y_{3}+K^{2}{y_{1}}^{2}\, \\   
\label{dy1}   
\frac{dy_{1}}{dl}\, &=&\,(1-K)y_{1}-4K^{2}y_{1}y_{3}\, \\   
\label{du}   
\frac{du}{dl}\, &=&\,-{\frac{1}{2}}uy_{1}^{2}(1+K){K}^{2}.   
\end{eqnarray}   
under a change of length scale $a \rightarrow a e^{dl}$. Here we define $ dl =  
\ln {\frac{a+da}{a}}$. The RG equation for the spin velocity $u$ is a 
consequence of the anisotropy of the $g_1$-interaction in the classical 2D 
XY-model, i.e. its non-retarded but non-local character in the quantum 
field theory (\ref{h1}).   
 
\begin{figure}
\centerline{\epsfysize=6.0cm \epsffile{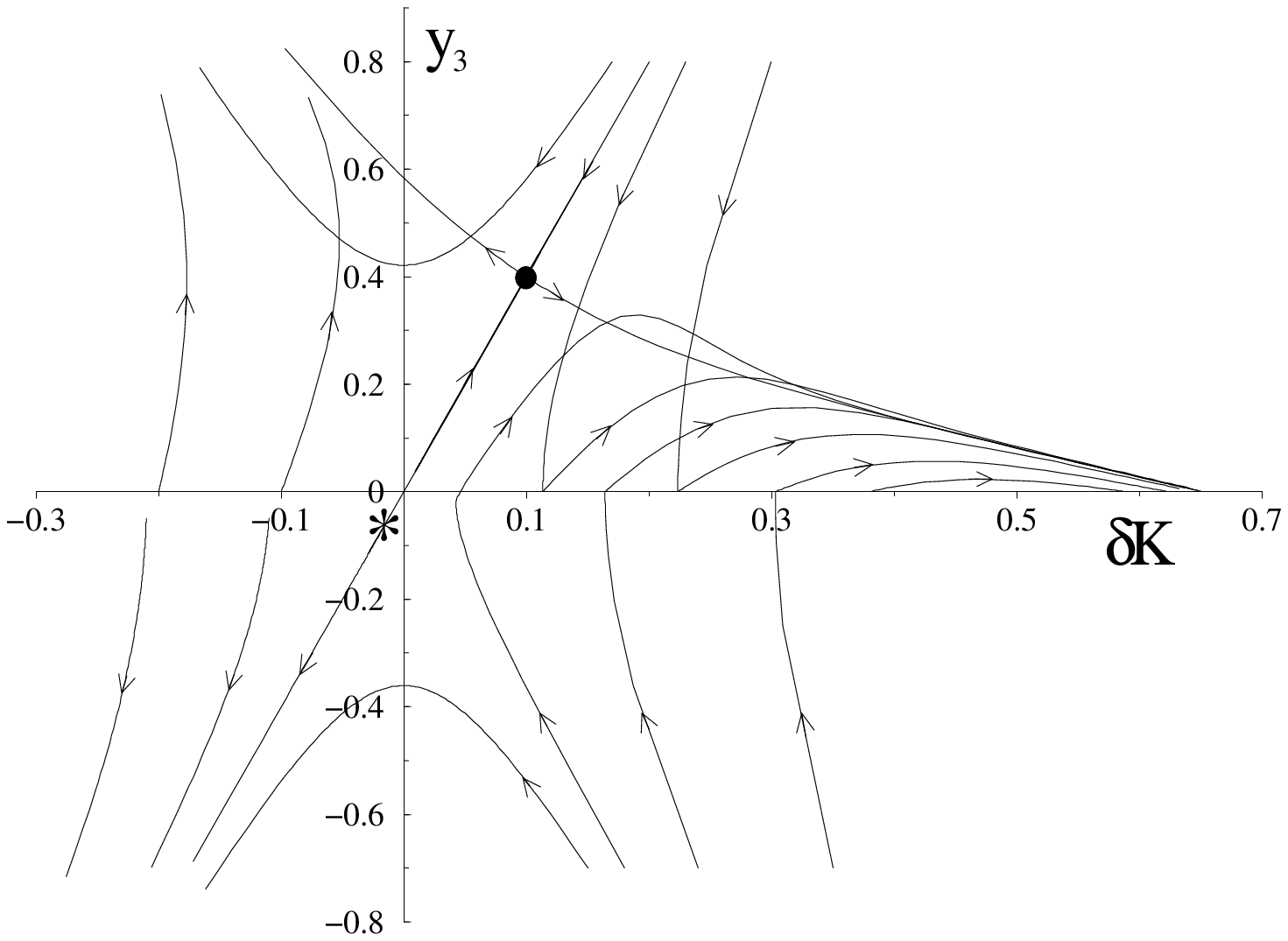}} 
\caption{The scaling trajectories for $y_1( l =0) =0.001$  
projected on the $y_3 - \delta K$ plane. $\delta K = K-1/2$, and the dot  
locates the new intermediate coupling fixed point. The N\'{e}el state is 
realized in the upper left, the dimer state in the lower left, and the  
spin liquid in the right part of the figure. The asterisk locates the boundary 
between flows to the new fixed point, and into the dimer regime.} 
\label{rg-fig} 
\end{figure}

For $y_1=0$ equations (\ref{dy1}) and (\ref{du}) would not appear, thus $u$ 
is not renormalized. The RG equations (\ref{dk}) and (\ref{dy3}) with $y_1=0$ 
describe the symmetric spin ladder system with a Kosterlitz-Thouless 
transition.\cite{Haldane} Spin-rotation invariant models scale along the 
separatrix between N\'{e}el
and spin liquid phases, or along its continuation into the dimer regime.  
Linearizing the RG equations (\ref{dk})-(\ref{du}) around the isotropic 
Heisenberg fixed point ($g_3^{(H)}=0, \; g_1^{(H)}=0, \; K^{(H)} = 1/2$) and 
defining $ \delta K = K -  \frac{1}{2}$, we obtain the linearized RG equations 
\begin{eqnarray}    
\frac{d~\delta K}{dl} \, & = & \,   
 - \frac{1}{4} y_3^2  + \frac{1}{16}y_1^2 \, , \\  
\frac{d~y_3}{dl} \, & = &  \,   
      - 4 \delta K y_3 + \frac{1}{4} {y_1}^{2} \, , \\  
\frac{d~y_1}{dl} \, & = & \,   
     \frac{1}{2}y_1 -  y_1 y_3 -\delta K y_1\, .  
\end{eqnarray}  
A family of solutions of the RG equations, 
projected on the $y_3-K$-plane, are shown in Fig. \ref{rg-fig}. We 
choose the initial value of $y_1$ as $0.001$ and find the trajectories are 
not sensitive to the choice of the initial values of $y_1$. From the RG 
equations, one can directly find that there exist two intermediate fixed 
points given by  $(\delta K,y_{3},y_{1})=(0.1,0.4,\pm0.8)$. Here $y_1$ takes 
the values $\pm$, which reflects that our RG equations is symmetric to $y_1$. 
As shown in Fig. \ref{rg-fig},  
the intermediate fixed point $(\delta K^{\star},y^{\star}_{3})=(0.1,0.4)$   
on the plane of $y_3 - \delta K$ is stable along the line  
$y_3 = 4 \delta K$  where  the spin-rotation invariance is protected.  
 For points near the intermediate fixed points, but not  
exactly on the line of $y_3 = 4 \delta K$,  the spin-rotation invariance is broken, thus  
they will flow to the spin-fluid fixed point $(y_3 =0)$ or the strong coupling  
fixed point $(y_3 \rightarrow \infty)$. 
 
At the fixed point, the RG equation (\ref{du}) implies that the spin velocity 
$u^{\star}$ is renormalized.  The robustness of the existence of this intermediate 
fixed point against higher order perturbations does not depend on the exact value of 
the fixed point as long as the fixed point is located on the RG separatrix with  
$1/2 < K < \infty$.\cite{CBV} 
The intermediate fixed point is thus described by an effective fixed-point Hamiltonian.
Inverting the definitions of    
$y_i = g_i / \pi u$, we rewrite the effective fixed-point Hamiltonian as   
a product of $u^{\star}$ and a term independent of $u^{\star}$,    
$H^{\star} = u^{\star} {\cal H}( K^{\star}, g_3^{\star}, g_1^{\star} )$.    
Then the vanishing of $u^{\star}$ leads to a trivial fixed   
point Hamiltonian $H^{\star} = 0$. As a result of the vanishing of the 
renormalized spin velocity, the elementary excitations at the fixed point, spinon 
and antispinon, are still gapless.  The  vanishing of velocity is usually 
interpreted as a sign of ferromagnetism, however we interpret this as our spins effectively 
decoupling at 
the lowest energy scales, i.e. a kind of asymptotic freedom in this
spin-rotation invariant ladder\cite{CBV}. Also, the numerical results of Wiessner et
al.\cite{Wiessner} indicate a paramagnetic susceptibility.
   
When $J_2$ increases beyond a critical value $J_{2c}(\delta)$ (now   
depending on $\delta$), the RG flows to a strong coupling fixed    
point, which corresponds to the quantum dimer phase. For $y_1 = 0.001$, this 
critical point is indicated in $(K,y_3)$-coordinates in Fig. \ref{rg-fig} by
an asterisk. For small $\delta$ and $J_2 > J_{2c}(\delta)$, our RG equations show 
that the system will remain in the universality class of the dimer solid 
corresponding to the strongly fixed point, however the spin gap is decreased 
by increasing $\delta$. Basically, the gap size $\Delta \propto \exp( -l_1)$ 
where $l_1$ is the length of the scaling trajectory from the initial values to 
the point where the most relevant perturbation is of order unity. This length 
is increased, and $\Delta$ therefore decreased, by the $y_1$ contributions to 
$K$ and $y_3$ being opposite in sign to those of $K$ and $y_3$ in Eqs.\ 
(\ref{dk}) and (\ref{dy3}).       
 
Recently,  Sarkar and Sen \cite{Sarkar} studied the same model by using a nonlinear 
$\sigma$-model field theory and Abelian bosonization. 
However, the main discrepancy between our
work and theirs \cite{Sarkar} is that we kept the bosonized  operator of
the alternating NNN operator and 
analyzed it by RG, while they just discarded it by giving an argument of
the irrelevance of the operator. For an anisotropic XXZ chain, our RG 
result indeed shows that  the $g_1$ operator is
irrelevant \cite{CBV} in the 
meaning that it does not drive the system to a new phase, and this is 
consistent with that of Sarkar et. al.  But the
main difference lies in the question whether an intermediate  fixed point exists 
and whether this fixed point corresponds to a phase different from a Luttinger liquid. In our
previous work the argument of a vanishing spin-wave velocity plays a crucial role
in the existence of such an unusual phase.  If the spin-velocity does not vanish, one should
explain the fixed point as a spin-liquid phase  as in the $J_1-J_2$ model with 
the spin velocity renormalized.  Our result also suggests that the quantum
phase transition parameter ($J_{2c} /J_1 \simeq 0.2412$) to the dimer phase is changed by 
the alternating NNN operator, which may be verified directly by numerical simulations like 
the density-matrix renormalization group. 
We also notice that the magnetization curve of the Heisenberg model with an
additional alternating NNN operator \cite{Wiessner} gives an obvious different
magnetization susceptibility from the one without it.
Their susceptibility is enhanced by this new interaction with respect to an equivalent 
Heisenberg chain, indicating a reduced spin velocity.  
Furthermore, a complete scheme to 
deal with the alternating NNN operator 
should give a correct description of the induced effect not only on the
weak frustration regime but also on the strong frustration regime.  The omission of 
the $g_1$ term could not give any explanation why 
the operator  shrinks the spin gap sizes in the regime
of strong frustration as we will study in detail in the next section. However,  our
RG analysis  gives a qualitative explanation of  the influence of the $g_1$ term on 
spin gap sizes. Therefore,  
we think that the scheme of discarding the $g_1$ operator based on its 
irrelevance seems to be oversimplified.  
We hope that more numerical simulations will eventually be able to 
resolve this disagreement and study quantitatively the phase diagram of the ground 
state as a function of $J_2/J_1$ and $\delta$. 
    
\section{Crossover from M-G to sawtooth model}   
It is generally believed that continuum field theory cannot give a good  
description for the behavior of the system far away from the critical point. 
In the case of $J_{2}=0.5J_1$, the correlations extend only to a distance 
of one lattice spacing, thus the continuum field description is not a good 
approach. As we have shown, the phase corresponding to large $J_2$ is the 
dimer phase. This is consistent with our knowledge from the models with 
exactly known ground states, say, Majumdar-Ghosh (M-G) model and sawtooth 
model. In this section, we prove that there exists a continuous manifold of 
Hamiltonians with dimer product ground states as long as $J_2 = J_1/2$.

We start with the asymmetric ladder model  
\begin{equation}   
\label{saw-mg}  
H=\sum_{l=1}^{2N} J  {\bf S}_{l} \cdot {\bf S}_{l+1} +  
 \left[ \frac{J}{2}+(-1)^{l+1} \delta \right] {\bf S}_{l} \cdot {\bf S}_{l+2}  
\; .      
\end{equation}  
The M-G model and sawtooth model are corresponding to $\delta=0$ and $\delta =  
J/2$ respectively.  
 
For the M-G model\cite{M-G}, the two linearly independent  
ground states, say, the left and right dimer ground states, are products   
of nearest-neighbor singlets, respectively   
\begin{equation}  
\label{mggstate}  
\mid \Phi _{L}\rangle =\prod_{l=odd}[l,l+1],~~\mid \Phi _{R}\rangle  
=\prod_{l=even}[l,l+1],  
\end{equation}  
where   
$ 
\lbrack i,j]= (\alpha _{i}\beta _{j}-\beta _{i}\alpha _{j}) / {\sqrt{2}}  
$ 
denotes the singlet combination of spin i and j with the direction of  
dimers defined as $i\rightarrow j$. Here $\alpha_i $ represents the up-spin and
$\beta_i $ the down-spin state at site $i$. $\mid \Phi _{L,R}\rangle  $ also  
represent the degenerate ground states of the   
sawtooth model\cite{Sen,Nakamura}. For the asymmetric ladder model, 
we notice that the NNN exchange alternation does not modify the product
states of nearest-neighbor singlets     
\begin{equation}   
H_{\delta }\mid \Phi _{L,R}\rangle = \sum_{l}   
(-1)^{l}\delta ~ {\bf S}_{l}\cdot{\bf S}_{l+2}\mid \Phi   
_{L,R}\rangle =0.   
\end{equation}   
This is induced by the fact that the alternating NNN couplings along the upper 
leg and the lower leg of the ladder cancel out each other, when they operate 
on $\mid \Phi_{L,R} \rangle$. It is obvious that $\mid \Phi_L \rangle$ and 
$\mid \Phi_R \rangle$ are eigenstates of the Hamiltonian (\ref{saw-mg}). 
In fact, as we will prove, they are exactly the ground states of (\ref{saw-mg}).  

To see this more clearly, we rewrite the asymmetric ladder model as a sum
of projection operators $P_{l}^{3/2}$:    
\begin{equation}   
\label{P-asylad}  
H=\sum_{n=1}^{N}  \left[ 
\frac{3}{2} ( \frac{J}{2} - \delta )  
P_{2n-1}^{3/2}  +  
\frac{3}{2} ( \frac{J}{2} + \delta )  
P_{2n}^{3/2} - \frac{3}{4}J \right] .     
\end{equation} 
with
\begin{equation}  
P_l^{3/2}= \frac{1}{3} \left[ \left({\bf S}_{l-1}+ {\bf S}_{l}+ {\bf S}_{l+1}  
\right)^{2} - \frac{3}{4} \right] .   
\end{equation}  
Here, we introduce $l$ to indicate the center position of three 
neighboring sites $(l-1,~l,~l+1)$. Such an operator is a special case of the general 
positive semidefinite  L\"{o}wdin's projection operators  \cite{Loewdin}  
\begin{equation}  
P^{S_{max}} = \prod_{S=S_{min}}^{S_{max}-1}\frac{\left({\bf S}_{1}+ {\bf  
S}_{2}+ \cdots + {\bf S}_{m} \right)^{2} - S(S+1)}{S_{max}(S_{max}+1)- S(S+1)}  
\end{equation}  
where $S_{max}$ and $S_{min}$ are the maximum and minimum values of the total  
spin $S$.  
  
As long as $|\delta| \leq \frac{J}{2}$, the coefficients in Eq.  
(\ref{P-asylad}) are non-negative. Therefore, the Hamiltonian (\ref{saw-mg}) is 
a linear combination of projection operators with positive coefficients. 
Since $P^{3/2}_l$ projects a state composed of three spins, $(S_{l-1}, 
S_{l}, S_{l+1})$, into a subspace of total spin $\frac{3}{2}$, its 
eigenvalues are $0$ (if the total spin is $\frac{1}{2}$) and  $1$ (if the 
total spin is $\frac{3}{2}$). By virtue of the properties of the positive 
semidefinite projection operator, whose lowest eigenvalue is zero, the ground 
state of the Hamiltonian (\ref{P-asylad}) can be constructed by choosing states 
with such configurations that each projection operator has the lowest 
eigenvalues $0$ when operating on these states. It is easy to prove that 
$\mid \Phi_{L,R} \rangle$ are the exact ground states of the asymmetric ladder model 
(\ref{saw-mg}). The ground state energy is independent of $\delta$ and given by 
\begin{equation} 
E_g = -\frac{3}{4}NJ . 
\end{equation} 
 
There is thus an entire manifold of Hamiltonians with fixed $J_1 = 2 J_2$,
parameterized by $\delta$, with doubly degenerate ground states of NN-dimer 
product ground states $\Phi_{L,R}$. For convenience, in our following
discussion we will shift our model by an energy of $E_g$, 
$
H - E_g \rightarrow H  \, ,  
$  
which is equivalent in taking the ground state of the system as zero.   

In the following, we consider the excited state of our system
(\ref{saw-mg}). The elementary excitation of the system is a pair of spinons
known as the kink or antikink\cite{Sen,Nakamura,CBV}.
First results were obtained by Shastry and Sutherland for the M-G
model\cite{S-S} within a variational ansatz. Since the ground state of
(\ref{saw-mg}) is independent of the alternating NNN exchange, the
construction of the excited states for the M-G model can be directly extended
to the asymmetric model (\ref{saw-mg}). Breaking a singlet pair in
the ground state would give rise to two unpaired ``defect'' spins. Therefore,
the simplest excitation consists of a pair of spinons. The spinons can be
thought as domain-walls separating different dimer ground state
configurations. From symmetry consideration, the kink and antikink are  
identical in the M-G  model, and specifically they have the same dispersions.
With alternating NNN interaction, the symmetry between legs is broken,   
therefore some properties of kinks and antikinks are different, in particular their
dispersion. However, they still survive as elementary excitations of  the
asymmetric spin ladder system.  
\begin{figure}
\centerline{\epsfysize=2.5cm \epsffile{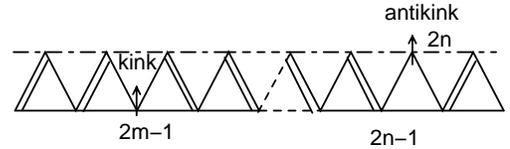}} 
\caption{{The kink and antikink excitations in the asymmetric ladder model.
The double lines represent singlets.}}  
\label{f3} 
\end{figure}  

In general, we call a spinon at the odd site $2m-1$ a kink and the other one at
the even site $2n$ an antikink (see Fig. \ref{f3}). The kink and
antikink always appear in pairs in a periodic system, however, a single spinon
can be realized in the open boundary systems. Taking the variational
wavefunction with one ``defect'' spin,\cite{S-S} one can easily obtain the
spinon dispersion of the M-G model
\begin{equation}
\varepsilon(k) = \frac{5}{8}J +\frac{J}{2} cos 2k.     
\end{equation}  
The energy gap $\Delta$ is therefore $J/8$. For $\delta = J/2$, i.e.\ the 
sawtooth model, the kink excitation is much different from the antikink 
excitation. As shown by D. Sen et. al. and T. Nakamura et. al. 
\cite{Sen,Nakamura}, the kink (K) excitation in the sawtooth chain is exactly
a single spin on odd site and dispersionless   
\begin{equation}  
\varepsilon_{K}(k) = 0.    
\end{equation}  
However, an antikink propagates with  an effective mass along the lattice.
The antikink is not a free spin and spreads out to an extended region,
because it is not a eigenstate of the local Hamiltonian. In the first
approximation, the antikink ($\overline K$) is supposed to be a single
defect spin at the even site, and the dispersion obtained by variational
calculation has similar form as the spinon dispersion of the M-G model    
\begin{equation}  
\varepsilon_{\overline K}(k) = \frac{5}{4}J + J cos 2k.   
\end{equation}  
with the corresponding energy gap of $J/4$. Despite the variational 
nature of the dispersion, the results agree very well with exact numerical 
results \cite{sorensen,Zheng}.   
 
We will explicitly calculate the change of the gap size with increasing
$\delta$. As expected, we found that the  $\delta$ term changes the energy gap
size of excitations, which is  consistent with our conclusion obtained by
the renormalization group analysis.  Following D. Sen et. al \cite{Sen}, we
assume both the kink and antikink to be a $5$-cluster block with spin $1/2$. It
is known that for the M-G and sawtooth chain there is no closely bound
kink-antikink pair whose energy is lower than that of a widely separated pair. 
Thus we can deal with the kink and antikink separately. The gap of the lowest
excitation is a sum of the gaps of kink and antikink
\begin{equation}   
\Delta ~=~ \Delta_K  ~+~ \Delta_{\overline K} ~,  
\end{equation}
where the subscripts, $K$ and $\overline K$, represent the
kink and the antikink respectively. It should be noticed that both the M-G
and sawtooth model have the same energy gap size $\Delta$ under the first 
approximation (1-cluster approximation). That is not true as we take 
more precise n-cluster approximation.      
 
\begin{figure}   
\centerline{\epsfysize=4.5 cm \epsffile{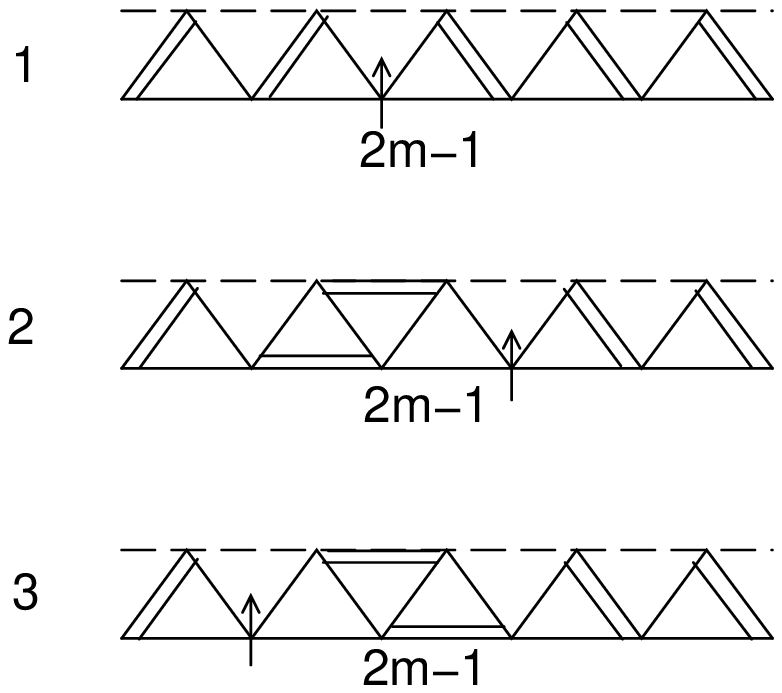} \epsfysize=4.5 cm  
\epsffile{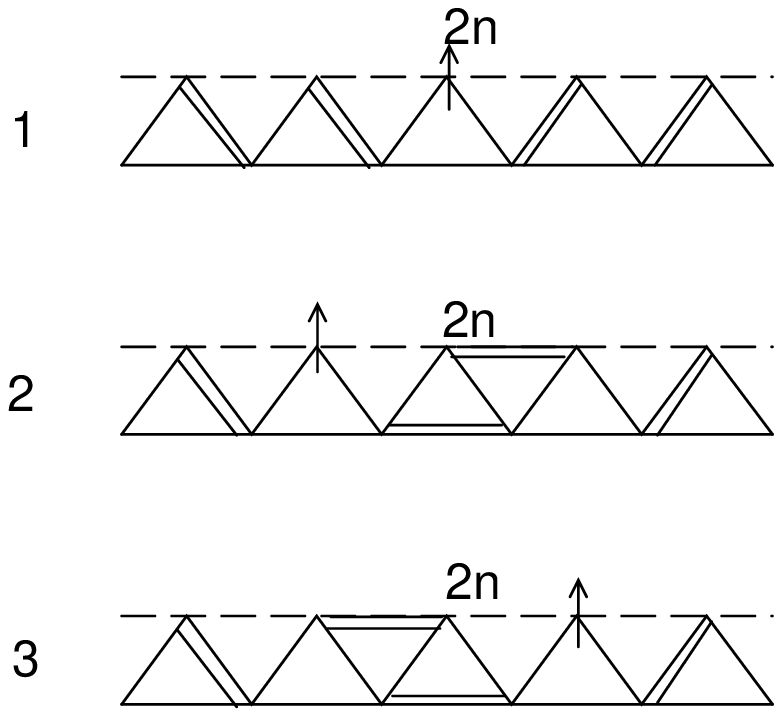}}  \caption{The five-size cluster of the kink and the
antikink.}    \label{cluster}    
\end{figure}

Under the $5$-cluster approximation, the only three linearly independent  
configurations that we need to consider are those shown in Fig.  
\ref{cluster}.  We denote these three configurations of kink by $\vert ~ 2m-1  
~\rangle_1 ~$, $\vert ~ 2m-1 ~\rangle_2 ~$ and $\vert ~ 2m-1 ~\rangle_3 ~$ and  
the configurations of antikink by $\vert ~ 2n ~\rangle_1 ~$, $\vert ~ 2n  
~\rangle_2 ~$ and $\vert ~ 2n ~\rangle_3 ~$ respectively. Here, $(2m-1)$ and $2n$  
denote the position of the center of 5-spin cluster corresponding to the kink and  
antikink. We now consider the momentum wavefunction with two variational  
parameters $a_{1,2}$ and $b_{1,2}$   
\begin{eqnarray}   
\vert k_{1} \rangle &=& {1 \over {\sqrt N}} \sum_m ~e^{i(2m-1)k_{1}} ~\left[  
\vert 2m-1 \rangle_1 + a_{1} \vert 2m-1\rangle_2  \right. \nonumber \\ 
 & & ~~~~~~ \left. ~+~ b_{1} \vert 2m-1\rangle_3  
\right]  \\   
\vert k_{2} \rangle &=& {1 \over {\sqrt N}} ~\sum_n ~e^{i2n k_{2}} ~\left[ ~  
\vert ~2n ~\rangle_1 + a_{2} \vert ~2n~\rangle_2  \right. \nonumber \\ 
& & ~~~~~~ \left. ~+~ b_{2} \vert ~2n~\rangle_3  
~ \right] ~,   
\end{eqnarray}  
where $k_{1}$ and $k_{2}$ are the momentum of the kink and antikink respectively.  
  
The lowest energy is obtained by finding parameters which minimize the  
energy expectation     
\begin{equation}  
\label{E_var}   
\varepsilon (k_{1,2})=\frac{\langle k_{1,2}|H|k_{1,2} \rangle }  
{\langle k_{1,2}|k_{1,2} \rangle }.    
\end{equation}   
Since $\vert 2m-1 \rangle_2$ and $\vert 2m-1 \rangle_3 $  
(~$\vert 2n \rangle_2 ~$ and $\vert 2n \rangle_3 ~$) are symmetric about the
site $2m-1$ ($2n$), there is no reason to discriminate between these
configurations and it is reasonable to choose $a_{1,2} = b_{1,2}~$.

The computation of Eq. (\ref{E_var}) is straightforward although a little bit 
lengthy, we will not give the detail here but refer to the literature  
\cite{Sen}. It is found that the minimum value of $\varepsilon (k_{1})$ occurs  
at $k_{1}= \pi/2$ and is given by
\begin{equation}  
\label{deltak}  
\varepsilon (k_{1} = \pi/2; a_{1}) ~=~ {1 \over 4} ~{{J/2~(1+4a_{1}^2) -
\delta } \over {1-a_{1}+a_{1}^2 /2}} ~,   
\end{equation}  
and the minimum value of $\varepsilon (k_{2})$ is  
\begin{equation}  
\label{deltaak}  
\varepsilon (k_{2} = \pi/2; a_{2}) ~=~ {1 \over 4} ~{{J/2~(1+4a_{2}^2) +
\delta}    
\over {1-a_{2}+a_{2}^2 /2}} ~,   
\end{equation}  
where we take $J=1$ for convenience.  
  
\begin{figure}   
\centerline{\epsfysize=5.cm \epsffile{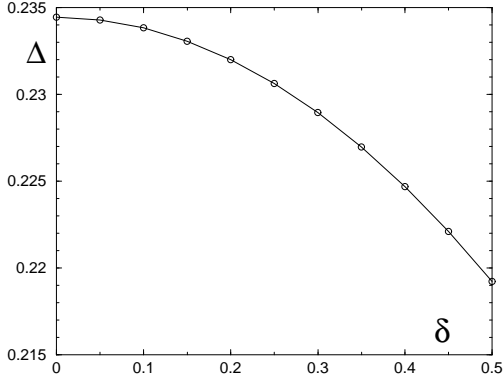}}   
\caption{The elementary excitation energy $\Delta$ versus $\delta$  
at the M-G point $J_{2}=0.5$ with $\Delta =  
\Delta_{K} + \Delta_{\overline K}$. }    
\label{exc}    
\end{figure}

For any given value of $\delta$, the excited gap $\Delta_K$  ($ 
\Delta_{\overline K}$) of a kink (an antikink) corresponds to the minimum 
of Eq. (\ref{deltak})  (Eq. (\ref{deltaak})). Our result is shown in Fig.
\ref{exc},  which indicates that the  elementary excitation gap decreases from
$0.234$ in the M-G to $0.219$ in the sawtooth with the increase of the
coupling constant  $\delta$, while the ground  state energy is constant.
In particular, for $\delta =  1/2 $, i.e. the sawtooth  chain,  Eq.
(\ref{deltak}) reduces to      
\begin{equation}  
\varepsilon (k_{1} = \pi/2; a_{1}) ~=~ {1 \over 4} ~{{2 a_{1}^2 }   
\over {1-a_{1}+a_{1}^2 /2}} ~,   
\end{equation}  
which has the minimum $ \Delta_K = 0$ for $a_{1}=0$, while Eq.  
(\ref{deltaak}) becomes
\begin{equation}    
\varepsilon (k_{2} = \pi/2; a_{2}) ~=~ {1 \over 4} ~{{1+2a_{2}^2}  
\over {1-a_{2}+a_{2}^2 /2}} ~.
\end{equation}  
which has a minimum $ \Delta_{\overline K}= 0.2192 ~$ at $a_2 = - 0.2808 $.   
It is clear that the kink excitation is exactly dispersionless, while an  
antikink is still a domain wall propagating with an effective  
mass\cite{Sen,Nakamura}.   
For the M-G model, $\delta = 0 $, Eq. (\ref{deltak}) and Eq. (\ref{deltaak})  
have the same form ($\Delta_K = \Delta_{\overline K}$), thus   
\begin{equation}  
\varepsilon (\pi/2; a)~=~ {1 \over 8} ~{{1+4a^2} \over  
{1-a+a^2 /2}} ~.   
\end{equation}  
whose minimum value is $0.2344 ~$. In these limits, the results are consistent  
with the known results of the M-G and the sawtooth model, as well as our
qualitative conclusion obtained from the field theory.  
  
If we introduce additionally an alternating NN exchange to our asymmetric
model  (\ref{saw-mg}), the degeneracy of $\mid \Phi _{L}\rangle $ 
or $\mid \Phi _{R}\rangle $ will be lifted
and the  singlets would be pinned along the stronger NN external dimer
potential.  The elementary excitations are no longer separated kinks and 
antikinks. In the presence of an external dimer potential, a kink and an 
antikink separated by a distance of $l$ give rise to a region in the 
incorrect ``ground state'', which effectively produces a confining potential
between the kink and antikink. This potential is proportional to the
distance of $l$, thus the kink and antikink can not escape from each other
and behave analogously as quark-antiquark pair. \cite{Affleck2,Dobry,Uhrig}
The kink-antikink bound state corresponds to a magnon with spin 1. The 
interesting topic of how the confined spinons develop to magnons has been 
investigated by Uhrig et. al.\cite{Uhrig2}
 
\section{Double layer model}
Recently, it was found that the 2D Shastry-Sutherland model\cite{Ueda,S-S2} 
can be used to explain the experimentally realized material $SrCu_2(BO_3)_2$,
hence such kind of model with exact dimer ground state\cite{CB} attracted much
attention again. In this section, we will show that the asymmetric spin 
ladder model can be generalized to a double layer model, whose ground state
is a simple direct product of singlet dimers.    

\begin{figure}   
\centerline{\epsfysize=5.cm \epsffile{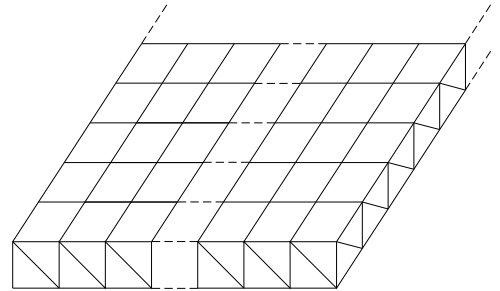}}   
\caption{The two-dimensional double layer model.}    
\label{layer-fig}    
\end{figure}

The double layer model is constructed from two coupled spin layers shown in Fig.
\ref{layer-fig}, where each layer has $N \times M$ sites and couples to the
other layer by the inter-layer exchange interactions $J_{\perp}$ and $J_{d}$.
The intra-layer exchange interactions $J^{1}$ and $J^{2}$ on top and bottom layers
may have different strengths. The Hamiltonian of our model is given by  
\begin{eqnarray}   
\label{layer2}  
H & = & \sum_{i,j=1}^{N, M} \sum_{\alpha=1}^{2} J^{\alpha} ( 
{\bf S}_{i,j} ^{\alpha} \cdot {\bf S}_{i,j+1}^{\alpha} +  {\bf S}_{i,j} 
^{\alpha} \cdot {\bf S}_{i+1,j}^{\alpha} ) \nonumber \\       
&+& \sum_{i,j=1}^{N, M} J_{d} \left( {\bf S}_{i,j} ^{1} \cdot {\bf 
S}_{i,j+1}^{2}  + {\bf S}_{i,j} ^{1} \cdot {\bf S}_{i+1,j}^{2} \right) 
\nonumber \\    
&+& \sum_{i,j=1}^{N, M} J_{\perp} {\bf S}_{i,j} ^{1} \cdot {\bf 
S}_{i,j}^{2} \; ,        
\end{eqnarray}  
where the superscripts $\alpha=1,2$ denote labels of the top and bottom
layers.  $J_{\perp}$ is the perpendicular inter-layer exchange interaction and
$J_{d}$ is the diagonal inter-layer exchange interaction. Here all the exchanges are 
taken to be positive. A similar model has been investigated in
Ref. \cite{Lin}, where the layer model is a direct generalization of
the Bose-Gayren ladder model\cite{BG}.
It is clear that every slice of the double layer net is just a ladder
whose Hamiltonian has the same form of Eq. (\ref{saw-mg}). Thus we find that
the ground state of the layer model is given by a product of all perpendicular
singlet pairs 
\begin{equation} 
\label{layer-dimer}  
\Phi _{D} = \prod_{i,j=1}^{M,N} \frac{1}{\sqrt{2}} 
(\alpha_{i,j}^{1} \beta_{i,j}^{2} -\beta_{i,j}^{1} \alpha_{i,j}^{2}) , 
\end{equation}
when the condition
\begin{equation} 
 J_{\perp} = 2J_{d}  =  2(J^{1} + J^{2}) 
\label{constraint}
\end{equation} 
is fulfilled. A rigorous proof can be made directly by representing the 
layer model as a sum of the projection operators as in the spin ladder case.
The corresponding ground state energy is  
\begin{equation} 
E_g = -\frac{3}{4}N M J_{\perp}  . 
\end{equation}  

It is obvious that the dimer product state $\Phi _{D}$ remains the ground state 
for $J_\perp > 2J_{d}$. The properties of the ground state are
independent of the specific values of $J^{1}$ and $J^{2}$ as long as the
constraint condition (\ref{constraint}) is satisfied. Since the dimerized
ground state is not degenerate, the lowest excitation is expected
to be a triplet excitation, corresponding to breaking of a singlet bond, with a gap 
size proportional to $J_\perp$.   However, the many-particle excitation spectra might 
be very complicated because of the effective interactions among the triplet 
excitations\cite{Kotov}.

\section{Conclusions}  
The spin-isotropic, asymmetric zig-zag ladders are
studied using the field theory method and the variational approach. When
the leg exchange integrals  are small compared with the NN exchange, the spin
model is mapped to a  revised double frequency sine-Gorden model.
Renormalization group analysis  shows that there are two fixed points, say, an
intermediate-coupling fixed  point and a strong coupling fixed point. In the
weak frustration limit,  the system is described by the
intermediate-coupling fixed point with gapless excitations. The vanishing of
spin velocity at the intermediate-coupling fixed point is likely to indicate a
decoupling of spins at low energy scales. Apart from the isotropic
separatrix, we find gapless spin liquid and gapped N\'{e}el states with
easy-plane and easy-axis anisotropy. For large frustration, a more usual
dimer solid phase is realized corresponding to the strong coupling fixed
point. The RG analysis also predicts that the spin gap is  decreased by
increasing the leg-asymmetry $\delta$. A continuous manifold of Hamiltonians
with the same singlet product ground state interpolates between the
Majumdar-Ghosh model and the sawtooth spin chain. Starting from the exact 
ground state wavefunction, we construct the variational wavefunction of the 
excited state and investigate the change of spin gap with the change of 
leg-asymmetry $\delta$. In the spirit of constructing Hamiltonian in the form
of a sum of positive semidefinite projection operators, extension to the
double layer model is carried out. We propose an exactly solved
two-dimensional double layer model with a ground state of a product of
inter-layer dimers.

\section*{Acknowledgments}
The authors would like to thank Prof. K. H. M\"{u}tter and Dr. M. Nakamura
for interesting discussions. S. Chen would like to specially thank F. Siano and 
B. Han for their critical reading of the manuscript. This research was supported 
by Deutsche Forschungsgemeinschaft through grants no. VO436/6-2 and VO436/7-2. 
 
\bibliographystyle{unsrt}  
 
\end{multicols}  
\end{document}